\magnification=\magstep1
\hsize=6.5truein  
\vsize=9.0truein  
\parindent=20pt    
\parskip=10pt      
\baselineskip=15pt 

\def\mone{$^{-1}$}
\def\mtwo{$^{-2}$}

\def\Msun{$M_\odot$}
\def\Lsun{$L_\odot$}

\def\deg{$^\circ$}

\def\esc{erg sec\mone\ cm\mtwo}

\def\\{\hfil\break}
\def\newpage{\vfill\eject}

\def\ein{{\it Einstein}}
\def\sig{$\sigma$}
\def\am{$'$}
\def\as{$''$}
\def\ein{{\sl Einstein}}
\def\about{$\sim$}
\def\Ha{$H\alpha$}

\def\chapter#1{\bigskip\noindent\begingroup\bigtype\bf #1 \endgroup\bigskip}
\def\section#1{\bigskip\noindent\begingroup\bf #1 \endgroup\medskip}
\def\subsection#1{\medskip\noindent {{\it #1}} \smallskip}

\def\approxlt{\mathrel{\hbox{\rlap{\lower 4pt \hbox {$\sim$}}$<$}}}
\def\approxgt{\mathrel{\hbox{\rlap{\lower 4pt \hbox {$\sim$}}$>$}}}
%
%  These provide a simple means of producing references in ApJ style.
%
\def\pp{\par\hangindent=.125truein \hangafter=1}
\def\apjref#1;#2;#3;#4{\pp #1, {\it #2}, {\bf #3}, #4.}
\def\ref#1;#2;#3;#4{\pp #1, {\it #2}, {\bf #3}, #4}
\def\book#1;#2;#3{\pp #1, {\it #2}, #3}
\def\rep#1;#2{\pp #1, #2}
%   \ref will construct a reference in Ap.J style.  The first 3 arguments
%           must be delimited by semicolons,  and the last by a space.
%           all four arguments must be supplied, e.g.
%        
%   \ref  Blow, Joe, 1917;J.I.R.;1;991
% 
\def\today{\ifcase\month\or
 January\or February\or March\or April\or May\or June\or
 July\or August\or September\or October\or November\or December\fi
 \space\number\day, \number\year}

%\input /home/kim/util/my_font.tex
%
%\nopagenumbers
%\hoffset=1truein
%\voffset=1truein

\magnification=\magstep1
\def\micron{$\mu$m}
\def\llam{$\lambda\lambda$ \/}
\def\lam{$\lambda$ \/}
\def\lamc{$\lambda_{c}$ \/}
\def\angst{{\rm \AA} \/}

\
\vskip 1truein

\centerline {\bf ROSAT X-RAY OBSERVATIONS OF}
\centerline {\bf THE RADIO GALAXY NGC 1316 (FORNAX A)}

\vskip 1.5truein
\bigskip
\centerline {D.-W. Kim}
\centerline {\it Chungnam National University, Taejon, 305-764, South
Korea; and}
\centerline {\it Harvard-Smithsonian Center for Astrophysics, 60
Garden Street, Cambridge, MA 02138}
\medskip\medskip
\centerline {G. Fabbiano and G. Mackie$^1$}
\centerline {\it Harvard-Smithsonian Center for Astrophysics, 60
Garden Street, Cambridge, MA 02138}
\vskip 1truein

\centerline {\today}

\vskip 1.5truein
\noindent {1. Visiting Astronomer, Cerro Tololo Inter-American Observatory.}

\newpage
\centerline {\bf Abstract}

We have observed NGC 1316 (Fornax A) with the ROSAT HRI. In this
paper, we present the results of these observations and we complement
them with the spectral analysis of the archival PSPC data. The
spectral properties suggest the presence of a significant component of
thermal X-ray emission ($>$ 60\%), amounting to $\sim 10^{9}$\Msun\ of
hot ISM. Within 3\am\ from the nucleus of NGC 1316, the HRI X-ray
surface brightness falls as $r^{-2}$. In the inner \about 40\as, the
X-ray surface brightness is significantly elongated ($\epsilon \sim$
0.3). This flattened X-ray feature is confirmed by a straightforward
statistics test as well as moment analysis.  By comparing the
morphology of the X-ray emission with the distribution of optical dust
patches, we find that the X-ray emission is significantly reduced at
the locations where the dust patches are more pronounced, indicating
that at least some of the X-ray photons are absorbed by the cold
ISM. We also compare the distribution of the hot and cold ISM with
that of the ionized gas, using recently obtained $H_\alpha$ CCD
data. We find that the ionized gas is distributed roughly along the
dust patches and follows the large scale X-ray distribution at r $>$
1\am\ from the nucleus. However, there is no one-to-one correspondence
between ionized gas and hot gas. Both morphological relations and
kinematics suggest different origins for hot and cold ISM. The radio
jets in projection appear to pass perpendicularly through the central
X-ray ellipsoid. Comparison of thermal and radio pressures suggests
that the radio jets are confined by the surrounding hot gaseous
medium.

\newpage
\centerline {\bf 1. INTRODUCTION}

NGC 1316 (Fornax A, Arp 154) is a giant elliptical galaxy in the poor
Fornax cluster. This galaxy exhibits many unusual features for an
elliptical galaxy, including pronounced dust patches, $H_\alpha$
filaments, ripples and loops (e.g., Arp 1966; Schweizer 1980; Carter
et al. 1983; Mackie and Fabbiano 1997). The distribution of the
optical surface brightness reveals an extensive envelope, making NGC
1316 a typical D (or cD) galaxy (Schweizer 1980).  It has one of the
most pronounced shell systems observed in early type galaxies (Malin
and Carter 1983).  These features all point to a recent merging in NGC
1316 (e.g., Schweizer 1980).

In the radio (0.03 - 5 GHz), Fornax A is one of the brightest objects
in the sky (L = $2 \times 10^{42}$ erg sec\mone; Ekers et al. 1983).
It contains giant radio lobes (Wade 1961), separated by \about 30\am\
(\about 240 kpc), consisting of polarized, organized filaments
(Fomalont et al. 1989). A faint bridge between the lobes is displaced
to the south of the galaxy center and S-shaped nuclear radio jets are
present, implying a significantly violent action in the galaxy
history, as also suggested by the optical data (e.g., Ekers et
al. 1983; Geldzahler and Fomalont 1984). The nucleus of NGC 1316 hosts
a low-luminosity AGN: optically it has a LINER-type spectrum
(Veron-Cetti and Veron 1986; Baum, Heckman and van Breugel 1992); it
contains a radio core (Geldzahler and Fomalont 1984); and HST
observations have revealed a nuclear UV-bright point source (Fabbiano,
Fassnacht and Trinchieri 1994b).  NGC 1316 also contains complex,
multiphase ISM: it has been detected in optical emission lines from
ionized gas (e.g., Schweizer 1980; Phillips et al.  1986; Veron-Cetti
and Veron 1986), IRAS far infrared emission (Knapp et al. 1989) and CO
lines of molecular gas (Wiklind and Henkel 1989; Sage and Galletta
1993).

In X-rays, NGC 1316 was detected with Einstein (Fabbiano, Kim and
Trinchieri 1992) and belongs to a group with the lowest $L_X / L_B$
ratio among E and S0 galaxies (Kim, Fabbiano and Trinchieri 1992b).
Therefore, based on the global amount of X-ray emission, 
NGC 1316 does not necessarily contain a large amount of hot
ISM (see Fabbiano, Gioia and Trinchieri 1988). In this, it is similar
to other galaxies which may have experienced recent mergers (Hibbard
et al. 1994; Fabbiano and Schweizer 1995). ROSAT PSPC (Feigelson et
al. 1995) and ASCA observations (Kaneda et al. 1995) have revealed the
presence of extended Inverse Compton X-ray emission at the locations
of radio lobes.

In this paper, we discuss the results of a re-analysis of the archival
ROSAT (Truemper 1983) PSPC observation of NGC 1316, and we report for
the first time the results of high resolution X-ray observation with
the ROSAT HRI.  The HRI has
\about 5 arcsec resolution (David et al 1993), comparable to that of
ground-based radio and optical data. With these data we establish the
presence of a hot ISM in NGC 1316 and we correlate its properties to
that of the other phases of the ISM. We also explore possible
interactions between this hot ISM and the active radio nucleus.

This paper is structured as follows: in section 2, we present the
results of the ROSAT HRI (\S 2.1 and \S 2.2) and the ROSAT PSPC data
analysis (\S 2.3); in section 3, we compare the X-ray data with the
optical and radio data; finally, in section 4 we discuss the
implications of our results.

\bigskip\bigskip
\centerline {\bf 2. X-RAY OBSERVATIONS}

NGC 1316 was observed with the ROSAT HRI on Jan. 14, 1994 and Jul.
7-10, 1994 for a total exposure time \about 40 ks (3/4 of total
observations were obtained in the July run). The two data sets are
consistent with each other within the observational uncertainty. We
present the combined results except as mentioned in \S 2.1. The field
of view of this observation also includes the companion elliptical
galaxy NGC 1317 (Figure 1). The observational log and basic parameters
of both galaxies are given in Table 1. NGC 1316 was also observed with
the ROSAT PSPC (Feigelson et al. 1995), and we used the PSPC archival
data to determine spectral parameters.

Given the limited statistics of our HRI observation, we have used
various binning and smoothing factors to investigate both large scale
(a few arcmin) and small scale ($\approxlt$ 1 arcmin) features. We
used both IRAF and software we developed ourselves to analyze our
data. We adopt a distance to NGC 1316 of 27.2 Mpc using $H_o$ = 50 km
s\mone\ Mpc\mone\ (Fabbiano et al. 1992).

To extract source data, the field background must be subtracted.  We
found that estimates from the background map generated by the standard
ROSAT SASS processing and from local measures of the background in
concentric annuli around the sources agree well with each other. We
will present results obtained with the local background subtraction.

\bigskip
\centerline {2.1 X-ray Emission Features and Sources}

The entire observed field of view is shown in Figure 1, where the
X-ray contour map is overlaid on the optical image obtained from the
Digitized Optical Survey$^{(2)}$. The X-ray image was binned with a
pixel size of 8\as\ and smoothed with a Gaussian of $\sigma$ =
16\as. The background is not subtracted. The octagonal shape indicates
the boundary of the HRI detector. The figure shows a strong X-ray
source at the center of the field, corresponding to the optical
position of NGC 1316.  Also X-ray emission is detected at the optical
position of NGC 1317 (6.3\as\ to north of NGC 1316).  Additionally, 10
point-like sources (above 3\sig) are detected in the observed
field. The source position, radius of the count extraction circle and
X-ray count for each source are listed in order of RA in Table 2. The
corresponding source numbers are marked in Figure 1. X-ray counts were
extracted from circular regions centered on their X-ray centroids. The
radius was determined with the radial profile of the X-ray surface
brightness (\S2.2.). Typically these radii extend to where the surface
brightness is within \about 3-4 \% of the background. The background
counts were extracted in annuli r=60\as\ - 200\as, except for NGC 1316
where an annulus r=200\as\ - 400\as\ is used.

\noindent -------------------------------------- \\
2. The Digitized Sky Survey was produced at the Space Telescope
Science Institute under U.S. Government grant NAG W-2166. The images
of these surveys are based on photographic data obtained using the
Oschin Schmidt Telescope on Palomar Mountain and the UK Schmidt
Telescope. \\
--------------------------------------

With the HRI observation of NGC 1316 we can trace the X-ray emission
out to 170\as, or 22.4 kpc at an adopted distance of 27.2 Mpc (see \S
2.2). However, the X-ray emission is extended at least to 1000\as\
(Fabbiano et al. 1992; Feigelson et al. 1995). The other sources
(including NGC 1317) are point-like.  Except for NGC 1316 and 1317,
the sources are not identified with known objects in the SIMBAD
catalog.

Source 2 and 12 are possibly variable. Their count rates from the
two HRI observing epochs (Table 1) are significantly different (above 3
\sig), while count rates of other sources are consistent within the
count errors between the two observations (see Table 3). The small
numbers of source counts do not warrant further temporal analysis.

The number of serendipitous sources (omitting the target galaxy) found
in this observation is consistent with the expected number of
background sources.  According to the LogN-LogS function (Hasinger et
al. 1993), we would have about 9 sources within a circular area of
radius 15\am\ and a limiting flux $1.8 \times 10^{-14}$ \esc\ in
0.5-2.0 keV, which corresponds to the faintest among detected sources
(Table 2) for a power law spectrum with $\alpha_E = 1$.

Figure 2 provides a close-up view of the NGC 1316 X-ray image which
was binned with a pixel size of 2\as\ and smoothed with a Gaussian of
\sig\ = 4\as, corresponding to FWHM = 11.3\as\ for an on-axis point
source. This image covers a $3' \times 3'$ field which falls inside
the optical galaxy ($12' \times 8.5'$ in $D_{25} \times d_{25}$; see
Table 1). Inside the central arcmin, the X-ray emission is elongated
along the NE-SW direction, following the optical major axis (PA=50\deg
; RC3), while in a larger scale (\about 2\am), the X-ray emission is
extended along the N-S direction. This extension is more pronounced
toward the north (see also Figure 1).

The observed NE-SW elongation inside r $\sim$ 1\am\ is intriguing if
real because it may be related to a disk forming in the rotating
cooling flow (\S 4.2) and because the radio jet in projection appears
to propagate perpendicular to the direction of this elongation (\S
4.3).  To determine its statistical significance, we first applied a
straightforward test by comparing counts and errors in angular sectors
at different position angles. We then applied more sophisticated
tests: a Monte Carlo simulation and moment analysis. Figure 3 compares
the counts at different position angles and gives the significance of
differences. Overall the difference between counts extracted in the
NE-SW (PA=30-70\deg, 210-250\deg) and NW-SE (PA=120-160\deg,
300-340\deg) is very significant with a signal to noise ratio of 7.7
$\sigma$ within r~$<$~40\as.

To estimate whether the flattening of the X-ray isophotes at small
radii is due to chance positioning of a few noise `blobs' from an
underlying spherical distribution, we have run a Monte Carlo
simulation. This involves: (1) choosing a spherical model; (2)
creating a set of `observations' in which Poisson noise is added to
the model and each X-ray is distributed taking into account the HRI
PRF; and (3) then determining the chance probability of occurrence of
the observed features. We generated a smooth model image by adopting
the X-ray radial profile determined in \S 2.2 under the assumption of
spherical symmetry. Then we normalized the model image and added
background counts to match with our HRI observation. We produced a set
of 100 `observations' with a pixel size of 2\as\ (as in Figure 2) by
estimating Poisson deviate for each pixel value and distributing the
photon according to the PRF. To estimate how often we obtain an
elliptical surface brightness by chance, we examined simulated images
after smoothing with the same Gaussian \sig\ as in Figure 2. Out of
100 images, no ellipticity as significant as that of Figure 2 is
seen. In our simulations, we also added serendipitous sources at
random positions according to the LogN-LogS function (Hasinger et
al. 1993) to verify whether a few undetected point sources can mimic
our observed features. However the results do not change
appreciably. In conclusion, the chance probability to generate the
observed, elongated distribution of the inner X-ray surface brightness
out of a circularly symmetrical distribution is less than 1\%.

To parameterize the X-ray surface brightness distribution, we have
applied a moment analysis (see fore details, Buote and Canizares 1994
and Carter and Metcalfe 1980). Ellipticities ($\epsilon$) and position
angles of the semi-major axis ($\theta$) are iteratively determined by
computing two-dimensional moments of inertia within elliptical
apertures. We used an unsmoothed, raw image (2\as\ pixel) with
background included. The measured ellipse parameters with a semi-major
axis varying from 10\as\ to 50\as\ are $\epsilon$ = 0.28 $\pm$ 0.05
and $\theta$ = 56\deg\ $\pm$ 8\deg. We have also applied ellipse
fitting (e.g., Jedrzejewski 1987), using the IRAF/STSDAS package to
the smoothed image in Figure 2.  The derived position angle and
ellipticity are consistent with those determined by the moment
analysis within their uncertainties.

Figure 4 shows the central region in greater detail with a pixel size
of 1\as\ and smoothed with a Gaussian of \sig\ = 2\as\ which
corresponds to an on-axis beam of 6.9\as\ FWHM. In this scale the core
surface brightness divides into double peaks, separated by 7.3\as. The
optical center (RC3) is close (within arcsec) to the X-ray centroid
determined with the image in Figure 2 and falls in the middle of the
double peaks shown in Figure 4. Also noticeable are larger scale
valleys at PA $\simeq$ 120\deg\ and PA $\simeq$ 320\deg\ which
apparently bisect the core if extended to the center.

The double peaks and the SE-NW X-ray valleys are not caused by
telescope aspect problems. We used only data obtained in the July run
(see Table 1), to avoid a possible mismatch of the galaxy centers in
the two observations due to aspect uncertainties, although source
positions differ by less than 2\as. The data obtained in the January
run also present the same features but with larger statistical
noise. To further check potential aspect problems, we applied the same
binning and smoothing to point-like sources within the observed
field. Sources 4, 7, and 8 which are within 5\am\ from the field
center [therefore the HRI PRF is similar to the on-axis one (David et
al. 1993)] are all circularly shaped while sources 2 and 12, both at
11\am\ off-axis distance (the PRF may not be circular), exhibit
randomly elongated distribution.

The observed features are very interesting, because the radio jet
(Geldzahler and Fomalont 1984) in projection appears to pass through
the X-ray valleys (\S 3). However, the relatively low X-ray counts
obtained in this high resolution image may produce an artificial
feature by chance positioning of a few noise `blobs'. To explore this
possibility, we have rerun a Monte Carlo simulation as described
above. Out of 100 simulations with a resolution same as Figure 4
(1\as\ pixel and \sig=2\as), 3 images exhibit double peaks and
valleys, although not as significant as the observed.  This simulation
results imply that the existence of the double peaks and the X-ray
valleys, although suggestive, is not conclusive and it is required to
be confirmed by a deeper observation.

\bigskip
\centerline {2.2 Radial Profile}

Figure 5 shows the radial profile of the X-ray surface brightness
measured in concentric rings centered on the NGC 1316 X-ray centroid
(Table 2). The raw, background and net counts are indicated by open
squares, a solid line and filled squares, respectively. The emission
is extended out to 170\as, 22 kpc at the adopted distance of 27.2 Mpc.
To this profile, we fitted a King approximation model, $\Sigma_X \sim
(1 + ({r \over a})^2 ) ^{-3 \beta + 0.5},$ convolved with the HRI
point response function. [In the case of gaseous emission, $\beta$ is
related to a true isothermal value by $\beta_i = 1.5 \times \beta$
(e.g., Sarazin 1988)].  Best-fit model, fit residuals, and confidence
contours are shown in Figure 6.  The best fit parameters and 90\%
confidence range with 2 interesting parameters (in parentheses) are:
core radius a = 4.1\as\ (3.1--5.0\as) and slope $\beta$ = 0.51
(0.49--0.54), yielding $\chi^2$ = 14.8 for 9 degrees of freedom. The
estimated slope corresponds to $\Sigma_X \sim r^{-2.06 \,
(1.94-2.24)}$ for $r >> a$.  This is close to that of the optical
brightness distribution. Using the data in Schweizer (1981), we
estimated the slope of the $V_4$ (5300-6400\AA) surface brightness in
r = 30\as\ to 300\as\ to be 2.02 $\pm$ 0.06. The radial slope measured
with the red continuum image (\S 3) does not differ sinificantly. We
also used the PSPC data to derive the radial profile of the X-ray
surface brightness. Using the radial profile within r=180\as\ (to
minimize the contribution of extended IC emission; see \S 2.3 and
Feigelson et al. 1995), and a background count rate estimated at
r=1800-2400\as\ (with vignetting correction), we find $\Sigma_X \sim
r^{-2.25 \, (2.12-2.45)}$, consistent with the HRI profile.

Fits to the radial profiles at the position angles of Figure 3, which
take into account the central flattening of the surface brightness
give slightly steeper slope [$\beta$ = 0.56 (0.52--0.61)] and slightly
larger core radius [a = 6.2\as\ (4.1--8.2)] along the direction of
flattening, whereas the opposite is true along the perpendicular
direction [$\beta$ = 0.46 (0.43--0.50) and a = 1.8\as\ ($<$ 3.8)].

A central point source that may be the X-ray counterpart of the radio
AGN was not detected with our HRI observations (see the HRI PRF
indicated as a dashed line in Figure 5; see also Figure 4). To
estimate an upper limit to a central point source, we applied the King
model plus the HRI point spread function in fitting the radial profile
and determined how strong a central source can be added without having
$\chi^2$ too large. We derive an upper limit (90\%) for the central
source of about 5\% of the total counts, which corresponds to a flux
of $1.0 \times 10^{-13}$ \esc\ in 0.1-2.4 keV and a luminosity of $9
\times 10^{39}$ erg sec\mone.  Here we assumed a
power law with an energy index $\alpha_{E}$=0.7 and line-of-sight
$N_H$. Of course a more luminous AGN would be allowed in the presence
of large intrinsic $N_H$.

The spectral analysis of the extended X-ray emission associated with
NGC 1316 (\S 2.3) suggests that over 60\% of the flux is likely to be
of gaseous origin. In the assumption that most of the extended
emission is due to a hot ISM, we have derived its 3-dimensional
density distribution, using a direct deprojection method (Kriss,
Cioffi and Canizares 1983), where the emissivity (or density) is
inwardly measured by subtracting the contribution of successive
spherical layers. We implicitly assume that the hot gas is homogeneous
and that the physical status of the gas at a given radius can be
represented by one temperature and density. The deprojected density
profile is shown in Figure 7a. We have assumed T = 0.8 keV (\S
2.3). The density profile corresponding to $\beta$=0.51, i.e., $n_e
\sim r^{-1.53}$ is also shown as a dashed line in the figure.

Using the deprojected density distribution, we estimate the cooling
time and gas pressure as a function of radius (Figure 7b). The cooling
time is given as $\tau_c = 1.5 \, {n k T \over n_e n_H \Lambda}$,
where $n, n_e$ and $n_H$ are the total particle density, electron
density, and Hydrogen density, respectively and $\Lambda$ is the
cooling function. The constant is in the range of 1 to 2.5, depending
on its definition (see Sarazin 1988).  The cooling time in the center
is \about $10^8$ years, much smaller than the Hubble time; it reaches
10$^{10}$ years at \about 180\as. To estimate the cooling function, we
assumed solar metal abundances. With the PSPC spectral data (\S 2.3),
the metal abundances cannot be determined unambiguously (see also
e.g., Trinchieri et al. 1994; Fabbiano et al. 1994a).  The cooling
time in the central region is still much shorter than the Hubble time
even with a zero metal abundance model. We also estimate the thermal
gas pressure using the measured density and assuming kT = 0.8 keV
(Figure 8).

We point out that these radial dependences of density, cooling time
and pressure are only indicative average values. Because of the
complexity of the surface brightness distribution in the inner
regions, we would expect a range of these physical parameters at each
radius, reflecting the clumpiness of the hot ISM. Also as remarked
earlier, we cannot study separately the properties of the different
components of the emission suggested by the PSPC data.

\bigskip
\centerline {2.3 ROSAT PSPC Spectral Properties}

In order to determine the X-ray spectral properties, we have used the
ROSAT PSPC data obtained from the public archives.  The method of data
reduction is similar to that of the PSPC observations of NGC 507 (Kim
and Fabbiano 1995). Because the X-ray emission is extended at least
out to 1000\as\ (see Feigelson et al. (1995) for discussions of the
extended structure of X-ray emission), we determined the background at
r=1800-2400\as\ and applied vignetting correction to each energy channel.
Since we are interested in the galaxy emission, we have analyzed the
X-ray emission within 180\as\ from the center.  Feigelson et al. (1995)
found Inverse Compton (IC) radiation at the location of the extended
radio lobes. The slope of the X-ray radial profile changes abruptly at
r\about 180\as, indicating that the X-ray emission in the central region
has a different origin from that of the extended X-ray emission
(Feigelson et al. 1995). To check for possible contribution of IC
radiation to the X-ray emission within r $<$ 180\as, we also estimated
the background using portions of the field within the PSPC support
structure (r \about 1000\as) where the diffuse emission is least.
However, the results are not significantly different.

Using XSPEC, we found that the X-ray spectra within 180\as\ can be
well reproduced either by a one-temperature, low abundance model (kT
$\simeq$ 0.6-0.7 keV and \about 10\% solar abundance) or by a
two-temperature, solar-abundance model (kT$_1$ $\simeq$ 0.1-0.2 keV
and kT$_2$ $\simeq$ 0.8-0.9 keV at 90\% confidence). In both cases,
the acceptable range of $N_H$ is consistent with the Galactic
line-of-sight value ($2 \times 10^{20}$ cm\mtwo). The best fit
parameters and acceptable ranges with $N_H$ fixed at the line-of-sight
value are listed in Table 4. Although we prefer the two-component
model rather than the almost zero-abundance model (see \S 5.1), both
models suggest a significant amount of hot gaseous emission, because a
significantly higher kT would be expected from a population of LMXB
(see the analysis of the M31 bulge data by Fabbiano, Trinchieri and
van Speybroeck 1987). We do not see any radial variations in the
spectral parameters (temperature and $N_H$) with the PSPC observations
(Table 4).

The absorption-corrected X-ray flux for the best fit parameters is $2
\times 10^{-12}$ \esc\ (see Table 5), corresponding to a X-ray
luminosity of $1.8 \times 10^{41}$ erg sec\mone\ at the adopted
distance of 27.2 Mpc. The HRI flux is consistent with the \ein\ IPC
flux (see Table 1).  For the two-component model, the X-ray flux of
the very soft component is \about 40\% of the total flux in 0.1-2.4
keV, similar to those seen in other X-ray faint early type galaxies
(Fabbiano et al. 1994a, Fabbiano and Schweizer 1995).  These results
indicate the presence of hot ISM contributing at least 60\% of the
total X-ray emission even if the soft component is fully of stellar
origin.

The analysis of the HRI image (\S 2.2) shows that a nuclear point
source cannot contribute significantly to the emission in the ROSAT
band. Therefore the spectral parameters (even in the outermost bin)
are representative of an extended emission component. To estimate the
upper limits of hard X-ray emission from low-mass binaries to this
emission [as seen in the bulges of spirals (Fabbiano 1989) and in
ellipticals (Matsushita et al. 1994; Matsumoto et al. 1997)], we added
a hard Bremsstrahlung component in the spectral fitting and determined
the acceptable range of its normalization. Its temperature was allowed
to vary between 3 and 20 keV.  The 90\% upper limit of the hard
component is 15\% of the total flux in 0.1-2.4 keV for a
one-temperature, low abundance model and 20\% for a two-temperature,
solar-abundance model. Even if we restrict our analysis to the central
1 arcmin region, the contribution of the hard component is less than
20\% for a 1-component model and 30\% for a 2-component model. This
upper limit corresponds to fluxes of $3-4 \times 10^{-13}$ \esc\ and
luminosities of $ 3-4 \times 10^{40}$ erg sec\mone\ (Table 5).  The
X-ray to optical luminosity ratio is then $L_X / L_B \le 2-3 \times
10^{29}$ erg sec\mone\
\Lsun\mone\ and is consistent with that of the bulge of M31 where the
X-ray emission is dominated by a population of individual bright X-ray
sources (e.g., Trinchieri and Fabbiano 1991) and with those of hard
components in other early type galaxies observed with ASCA (Matsumoto
et al. 1997).

\bigskip\bigskip
\centerline {\bf 3. COMPARISON WITH OTHER WAVELENGTH DATA}
\bigskip

Figures 9, 10 and 11 compare the high resolution HRI contour map of
Figure 2 with the distribution of dust (Schweizer 1980), ionized gas,
and radio continuum (Geldzahler and Fomalont 1984),
respectively. Although the low signal-to-ratio nature of the X-ray
data is such that a definitive, quantitative comparison can not be
done, there are some qualitative trends that it is worth to mention.

In general, the hot and cold ISM are related in the sense that the
X-ray emission is significantly reduced at the locations where the
dust patches are more pronounced. In particular, a dust lane is found
in the central SE X-ray valley, suggesting the possibility of
absorption.

Because the ionized gas may be originated from the cooling hot gas,
the distribution of ionized gas provides further clues of the
relationship between cold and hot ISM. Narrow-band CCD images were
taken on the photometric night of 1994 November 11/12.  The
CTIO/University of Michigan Curtis Schmidt 0.6/0.9m telescope was used
with a Thomson 1024x1024 CCD.  Pixel size is 19\micron\ square
(1.835\as) however vignetting limits the useable field size to about
30\am.  Total exposure times were 8100s each for the redshifted
(v=1801 km/s) \Ha\ + [NII] (\lam 6563 + \llam 6548,6583) emission line
filter (\lamc = 6606\angst, FWHM=76\angst), and continuum filter
(\lamc = 6693\angst, FWHM=81\angst). Bias and dark frames were
taken. Flatfields were generated from twilight sky exposures. The
spectrophotometric standard HZ 4 was used to calibrate the narrow band
images based on AB magnitudes (Oke 1994, private communication). The
adopted magnitudes were AB$_{6606}$=15.01 and AB$_{6693}$=14.81. The
\Ha\ + [NII] image was derived by subtracting a scaled, sky subtracted
continuum image from the sky subtracted emission line image. The
adopted scaling was calculated from a linear least square fit to
residuals of 25 field stars calculated from several scaling
factors. The mean pixel value of the \Ha\ + [NII] image at large radii
of the optical galaxy was also consistently near zero.  The
distribution of ionized gas is shown in Figure 10, superposed on the
X-ray image (same as Figure 2). The overall distribution of the
ionized gas is similar to that of the dust, ie., aligned toward
north-south, slightly turned to NW and SE (clockwise), as observed in
other early type galaxies with cold ISM (Kim 1989). The peak of the
northern blob ($40^s.5$, $11'.8$) falls in between the dust patches,
likely due to dust absorption. The eastern part of the southern blob
generally follows the distribution of the dust patch.  Comparing the
distributions of hot and warm ISM, we also find an overall similarity
(e.g., N-S extension), but there is not always a one-to-one
correspondence.

Figure 11a shows the radio jets superposed on the X-ray image (same as
Figure 2). This radio image was reproduced from the Figure 2b in
Geldzahler and Fomalont (1984), which was obtained with the VLA and
has 4\as\ resolution at 1.5 GHz. The direction of the radio jet is in
projection perpendicular to the direction of NE-SW elongation of the
central X-ray distribution. If confirmed, the X-ray valleys (seen in
Figure 4) may imply even more striking relations in that both sides of
the radio jets in projection coincide with the X-ray valleys at
PA$\simeq$120\deg\ and PA$\simeq$320\deg\ (Figure 11b), suggesting
that the radio jets are interacting with the surrounding hot gaseous
medium as seen in Cygnus A (Carilli, Perley and Harris 1994).

These connections need to be checked with future high resolution
observations. The AXAF CCD detector will provide 1\as\ resolution data
with spectral resolution as well, which will be ideally suited for
this type of investigation.

\bigskip\bigskip
\centerline {\bf 4. DISCUSSION}
\bigskip

\centerline {4.1 The Nature of X-ray Emission of NGC 1316}

The X-ray emission of E and S0 galaxies can be due to different
sources (see Fabbiano 1989): a hot gaseous halo dominating the
emission in the X-ray bright early type galaxies; integrated stellar
X-ray binary emission, seen in the bulges of spirals (see Fabbiano
1989) and confirmed in ellipticals with ASCA (Matsushita et al. 1994;
Matsumoto et al. 1997); a nuclear source, seen in bright radio
galaxies (Fabbiano et al. 1984; Worrall and Birkinshaw 1994); and a
very soft component, seen in X-ray-faint early type galaxies (Kim et
al. 1992b; Fabbiano et al. 1994a; Kim et al. 1996) of debatable nature
(e.g., Pellegrini and Fabbiano 1994). NGC 1316 is both an X-ray faint
D (maybe cD) galaxy and a radio galaxy, therefore its X-ray emission
is likely to be complex.

With the HRI imaging data and the PSPC spectral data, we can limit the
contribution from the nuclear component and stellar binary component
respectively as being relatively unimportant to the detected X-ray
emission (Table 5). The upper limit to the nuclear X-ray emission
obtained with the HRI image (\S 2.2) is $L_X
\le 9 \times 10^{39}$ erg sec\mone, which is about 5\% of the total
luminosity.  Of course a more luminous nuclear source could be present
and not visible if the nuclear $N_H$ is much higher than the line of
sight one. With the PSPC spectra, we can pose an upper limit to the
hard stellar component of 20-30\% of the total flux in 0.1-2.4 keV
($L_X \le 3-4 \times 10^{40}$ erg sec\mone; \S 2.3). This is
consistent with the expected hard X-ray emission from a 'bulge'
population of X-ray sources (e.g., scaling from the M31 bulge;
Trinchieri and Fabbiano 1991). Individual typical X-ray binary sources
would not be detectable at the luminosity threshold of our observation
(a few x $10^{39}$ erg sec\mone).

The PSPC spectrum of NGC 1316 (\S 2.3) suggests that most of the X-ray
emission within 3\am\ is due to a hot ISM in this galaxy.  Although
with the PSPC data we cannot unequivocally define the emission model,
the spectral fits require the presence of a \about 0.7-0.9 keV thermal
emission (Table 4). The X-ray spectrum of NGC 1316 resembles that of
other X-ray faint early type galaxies. X-ray spectra of those galaxies
observed both with the IPC (Kim et al. 1992a) and the PSPC (e.g.,
Fabbiano et al. 1994a) present an excess of counts in the lowest
energy channels, when compared to those of X-ray bright E and S0. This
type of spectrum can either be fitted with a metal-free
single-temperature optically-thin model, or with two (or more)
component models (see Fabbiano et al. 1994a; Pellegrini and Fabbiano
1994; Fabbiano and Schweizer 1995). Recent ASCA measurements of the
spectrum of the X-ray faint galaxy NGC 4382 (Kim et al. 1996) reject
the single-component metal-free model in that galaxy.  The two
component model, in the case of NGC 1316, would consist of a very soft
component of kT \about\ 0.1-0.2 keV (which could be of stellar origin,
e.g., Pellegrini and Fabbiano 1994), and a harder (kT \about\ 0.7-1.0
keV), almost solar metallicity ISM. It is interesting that the
abundance of the ISM determined from this two-component model is near
to the optical metallicity of this galaxy (e.g., Gorgas, Efstathiou
and Salamanca 1990). However, based on more general ASCA results on
galaxies (e.g., Loewenstein et al. 1994; Arimoto et al. 1996), the
latter may just be a coincidence.

Although a hot ISM is present, the total amount of hot gas is \about
$10^{9}$ \Msun, only a fraction of that present in X-ray bright early
type galaxies with a comparable optical luminosity (for example,
$M_{gas} \sim 2 \times 10^{10}$ \Msun\ in NGC 4636; Trinchieri et
al. 1994). NGC 1316 clearly has not been accumulating all the gas
ejected from the evolved stars during a Hubble time, which would be
\about $3 \times 10^{10}$ \Msun. This is not surprising since
this galaxy may have undergone recent merging events (see Schweizer
1980). Recent mergers (Hibbard et al.  1994) and dynamically young
ellipticals (Fabbiano and Schweizer 1995) also tend to be relatively
empty of hot ISM. It may be a coincidence, but the mass of the hot ISM
is comparable with what would be expected from stellar accumulation in
\about 1 Gyr, which may be the age of merger (Schweizer 1980). In a
recent paper discussing the optical and X-ray emission at larger
radii, Mackie and Fabbiano (1997) find
\about$3 \times 10^8$ \Msun\ of hot ISM spatially coincident with the
tidal tails in the outskirts of NGC 1316.

\bigskip
\centerline {4.2 Multi-phase ISM}

The comparison between X-ray and optical data reveals that the
distribution of the X-ray emission is related with that of dust
patches (\S 3.1.). The X-ray emission is weak where dust is seen and
X-ray blobs are often surrounded by dust patches. The X-ray valleys
running toward NW and SE from the center (Figure 4), if confirmed, may
be real low density regions in the hot ISM, because they are not
coincident with dust patches, except in the central SE region. The
ionized gas is overall cospatial with the dust patches. However, the
northern blob of the ionized gas falls in between dust patches, while
the southern blob coexists with dust, indicating some of the line
emission is absorbed by the dust.

The origin of the ionized gas (i.e., ionization mechanism) is unclear
in early type galaxies. The emission lines often indicate LINER-type
nuclear activity rather than HII regions, judged by for example the
relative line strengths \Ha\ to [NII] flux ratio (e.g., Kim 1989). It
has also been suggested that the gas is photo-ionized by post AGB
stars (Trinchieri and di Serego Alighieri 1991). In the case of
galaxies containing significant amounts of hot ISM, the ionized gas
may also be the result of cooling flows (e.g., Sarazin 1988), but this
is unlikely in NGC 1316.

In NGC 1316, the kinematics of the ionized gas revealed that the gas
is rapidly (up to 350 km sec\mone) rotating along the minor axis
(PA=127-142\deg ; Schweizer 1980), while the stellar system rotates
along the major axis (PA=50\deg ; Bosma, Smith and Wellington
1985). The same is true for molecular gas (Sage and Galletta
1993). This suggests an external origin for the cold and ionized ISM,
perhaps connected with the merging episode (Schweizer 1980; Mackie and
Fabbiano 1997) and argues against the idea that the ionized/cold gas
is originated from cooling hot gas. In the latter case both hot and
cold/warm gas would be expected to have the same kinematics as the
stars, since the hot ISM is likely to originate from stellar
evolution. Different origins of warm gas (likely cold gas and dust as
well) and hot gas may be further supported by our HRI observations in
that the central X-ray flattening, possibly a disk would
rotate along the major axis and that the morphological relationships
between the ionized gas and the X-ray emission are lacking. Therefore,
it is likely that the cold and warm ISM might be acquired externally
by mergers and infalls occurred \about 10$^9$ years ago, while the hot
gas has been accumulated since the latest merger.

%\bigskip
%\centerline {4.2 Disk formed in a rotating cooling flow?}

% We have found (\S 2.1) that the distribution of X-ray surface
% brightness is elongated (ellipticity \about\ 0.3) along the optical
% major axis (PA $\simeq$ 50\deg) within the central \about 40\as\
% (\about 4 kpc). 

If the infalling, cooling hot ISM carries angular momentum it may form
an accretion disk, which could be extending out to a 10~kpc radius
(see Kley and Mathews 1995; Brighenti and Mathews 1996).  The NE-SW
flattened isophotes of NGC 1316 (Figure 2) may represent such a disk.
However, because the isophotes are measured at radii less than the
optical effective radius, the stellar potential needs not be round at
these radii. Therefore, the observed elongation may be consistent with
hydrostatic gas in the stellar potential (or with a modest amount of
dark matter).

\bigskip
\centerline {4.3 Is the radio jet thermally confined?}

Interestingly, the observed elongation in the hot gas structure is
approximately perpendicular to the radio jet of NGC 1316. The
possibility of thermal confinement is supported by the radial
behaviour of the jet/ISM energetics. We estimated the radio jet
pressure corresponding to the minimum energy using the radio map by
Geldzahler and Fomalont (1984) (see Feigelson et al. 1995 for the
validity of the minimum energy argument). We applied the prescription
for a Gaussian jet given in Killeen, Bicknell and Ekers (1986a; see
also Pacholczyk 1970).  We also used the cylindrical jet approximation
(see Perley, Willis and Scott 1979), but the results do not change
significantly.  We assumed that: the jet and the magnetic field are
perpendicular to the line of sight; the energy of relativistic
electrons is equal to that of protons and ions; the radio spectrum is
a power law from 10 MHz to 10 GHz with a slope $\alpha$=0.7; and the
volume filling factor is unity.  The estimated jet pressure is
compared with the thermal pressure in Figure 8. At 6 - 24\as\ from the
center, the jet pressure for both NW and SE jets is an order of
magnitude lower than the thermal pressure. Radio and thermal pressure
are supposed to be in balance in the case of thermal
confinement. However, as discussed by Killeen et al. (1988) the radio
minimum pressure could be easily underestimated (see also Pacholczyk
1970) while the similarity of the radial behaviour of radio and X-ray
pressures argues for thermal confinement.  This apparent contradiction
between the thermal gas and minimum radio pressures has also been
reported in similar cases where radio jets are expanding through the
hot gaseous environment (e.g., Bohringer et al. 1993; Carilli, Perley
and Harris 1994).

The possibility (suggested by Figure 8) that the jet is thermally
confined inside a relatively small region reinforces the suggested
lack of causal connection between the jet and the extended lobes
(Ekers et al. 1983). It is possible that active events took place some
time ago, probably induced by a merger (\about 10$^9$ years ago;
Schweizer 1980) and now the nucleus is relatively weak and the radio
lobes are slowly cooling (see also Ekers et al. 1983). This idea is
also supported by the relative power (\about 1/2500) of the jets to
the extended radio lobes, which is a few hundred times lower than that
of a typical early type galaxy (Slee et al.  1994); the lack of strong
optical emission lines (see Schweizer 1980); the absence of a
connection between the jet and the lobe (Geldzahler and Fomalont
1984); and the morphology of the lobes (Fomalont et al. 1989).

The double peaks and X-ray valleys seen in Figure 4, if confirmed,
will provide more direct evidence for interaction between the hot gas
and the radio jet. Both sides of the radio jets in projection appear
to pass through the X-ray valleys which may play a role as nozzles in
collimating the jets.

\bigskip
\centerline {4.4 The Nucleus}

Fabbiano et al. (1984) found a relationship between radio core power
and X-ray luminosity in a sample of 3CR galaxies, indicating that both
radio and X-ray emission are of non-thermal nuclear origin.  This
relationship holds down to radio faint galaxies (Fabbiano et
al. 1989). Recently, with ROSAT observations, Worrall and Birkinshaw
(1994) spatially decomposed central X-ray point sources from the
diffuse, extended emission in several radio galaxies and confirmed the
linear relationship between core radio and central X-ray emission in
low-power radio galaxies.  Although the X-ray core of NGC 1316 is not
detected, the ratio between the X-ray upper limit (corresponding to
$l_{1 keV} \le 6.3 \times 10^{21}$ erg s\mone\ Hz\mone) and the
radio core emission ($l_{5 GHz} = 2 \times 10^{28}$ erg s\mone\
Hz\mone, form Geldzahler and Fomalont 1984) is consistent with the
linear relationship between these quantities discussed above.

A comparison of the spectral energy distribution (SED) of the NGC 1316
nucleus with those of other LINER galaxies can be found in Fabbiano
and Juda (1997). It is worth noting the possible similarity between
the nuclear sources of NGC 3998 and NGC 1316 (both early type
galaxies).  They both present a bright UV point-like source discovered
with HST, and their SED differ from those of bright AGN (see also
Fabbiano et al. 1994b). A better coverage of the nuclear emission of
these faint AGN will be necessary to understand the emission
mechanism.

\bigskip\bigskip
\centerline {\bf 5. CONCLUSION}
\bigskip

We have presented the analysis of the high spatial resolution image and
of the X-ray spectrum of NGC 1316 (Fornax A) obtained with the ROSAT
HRI and PSPC. The results lead to the following conclusions:

(1) The X-ray emission of NGC 1316 is extended. No point sources are
detected within the galaxy at a luminosity threshold of a few x
$10^{39}$ erg sec\mone. The radial profile of the X-ray surface
brightness falls as $r^{-2}$, which is close to the optical light
distribution.  No gradient of the X-ray emission temperature is seen.
Within the central 40\as, the X-ray isophotes are flattened along the
optical major axis. In a larger scale (1-2\am), the X-ray emission is
extended toward N-S, in agreement with the PSPC report of Inverse
Compton emission (Feigelson et al. 1995).

(2) The X-ray spectrum of NGC 1316 (r $<$ 180\as) can be reproduced
either by a single-temperature low-abundance model (kT = 0.7 keV and
10\% solar) or by a two-temperature, solar-abundance model (kT$_1$ =
0.1-0.2 keV and kT$_2$ = 0.8-0.9 keV). These results indicate the
presence of hot gaseous component contributing to $>$60\% of the total
X-ray emission. We set an upper limit of \about 20\% of the total
emission due to a hard component from LMXB in NGC 1316, consistent
with an extrapolation based on the bulge of M31 as well as those of
early type galaxies observed with ASCA.  The total X-ray emission is
$2
\times 10^{41}$ erg sec\mone\ in the  0.1-2.4 keV band and $M_{gas} \sim
10^9$\Msun.  The relatively small amount of hot ISM present in this
X-ray faint galaxy [by comparison with that of X-ray bright E and S0
such as NGC 4636 and NGC 4472 (e.g., Fabbiano et al. 1992)] is
consistent with observations of other systems that may have undergone
relatively recent merging (Fabbiano and Schweizer 1995).

(3) The X-ray emission is significantly reduced at the locations where
the dust patches are more pronounced, indicating that some of the
X-ray emission may be absorbed by the internal cold ISM in NGC
1316. The ionized gas is generally distributed along with the dust
patches. Some features of the ionized gas appear to be related with a
hot ISM but there is no one-to-one correspondence.  Both morphological
relations and kinematics suggest that the hot and cold/warm ISM may
not have the same origin. The cold/warm ISM might be acquired
externally by mergers/infalls, while the hot ISM has been accumulated
since the latest merger.

(4) In projection, the direction of the radio jets is perpendicular to
the central NW-SE elongation of the X-ray emission. The thermal
pressure is higher than the jet equipartition pressure by an order of
magnitude. However, the radial behaviour of thermal and radio
pressures are similar, suggesting the possibility of thermal
confinement.

(5) Although a nuclear source is not detected in X-rays, the upper
limit ($L_X \le 9 \times 10^{39}$ erg sec\mone) is consistent with the
expectations, based on the extrapolation from low-power radio galaxies
(Worrall and Birkinshaw 1994).

\bigskip\bigskip
\centerline {\bf ACKNOWLEDGMENTS}

\noindent This work was supported by NASA grants NAG 5-2152 (ROSAT),
NAGW 2681 (LTSA), NASA contract NAS8-39073 (AXAF Science Center) and
by Non Directed Research Fund, Korea Research Foundation. This
research has made use of SIMBAD which is operated by the Centre de
Donnees astronomiques de Strasbourg (CDS), France.  We thank the
referee and Greg Bothun for their comments on the original draft.

\newpage
\hsize=6.2 true in
\hoffset= 0.3 true in
\parindent= -0.3 true in
\parskip=0pt
\centerline {\bf REFERENCES}
\bigskip

Arimoto, N., Matsuchita, K., Ishimaru, Y., Ohashi, T., \& Renzini, A. 1996 preprint

Arp, H. 1966, ApJS, 14, 1

Baum, S. A., Heckman, T. M. \& van Breugel, W. 1992, ApJ, 389, 208

Begelman, M. C., Blandford, R. D., \& Rees, M. J. 1984, Rev. Mod. Phys. 56, 255

Bohringer, H., Voges, W., Fabian, A. C., Edge, A. C., \& Neumann, D. M. 1993, MNRAS, 264, L25

Bosma, A., Smith, R. M. \& Wellington, K.J. 1985, MNRAS, 212, 301

Brighenti, \& Mathews, W. 1996, preprint

Buote, D. A., \& Canizares, C. R. 1994, ApJ, 427, 86.

Carilli, C. L., Perley, R. A., \& Harris, D. E.  1994, MNRAS, 270, 173

Carter, D., Jorden, P.R., Thorne, D.J., Wall, J.V. \& Straede, J.C. 1983 MNRAS, 205, 377

Carter, D., and Metcalfe, N. 1980, MNRAS, 191, 325

David, L. P., Harnden, F. R., Kearns, K. E., \& Zombeck, M. V. 1993, The ROSAT High Resolution Imager (US ROSAT Science Data Center/SAO)

de Vaucouleurs, G., de Vaucouleurs, H., Corwin, H. G., Buta, R.  J., Paturel, G., \& Fouque, P. 1991 {\it Third Reference Catalogue of Bright Galaxies}, ({\bf RC3}), (New York: Springer-Verlag)

Ekers, R. D., Gross, W. M., Wellington, K. J., Bosma, A., Smith, R. M., \& Schweizer, F. 1983, AA, 127, 361

Fabbiano, G. 1989, ARAA, 27, 87

Fabbiano, G., Miller, L., Trinchieri, G., Longair, M., \& Elvis, M. 1984, ApJ, 277, 115

Fabbiano, G., Gioia, I. M., \& Trinchieri, G. 1988, ApJ,324,749

Fabbiano, G., Gioia, I. M., \& Trinchieri, G. 1989, ApJ, 347, 127

Fabbiano, G., Kim, D.-W., \& Trinchieri, G. 1992, ApJS, 80, 531

Fabbiano, G., Kim, D.-W., \& Trinchieri, G. 1994a, ApJ, 429, 94

Fabbiano, G., Fassnacht, C.,  \& Trinchieri, G., 1994b, ApJ, 434, 67

Fabbiano, G., \& Schweizer, F. 1995, ApJ, 447, 572

Fabbiano, G., \& Juda, G. 1997, in preparation.

Fabbiano, G., Trinchieri, G., and van Speybroeck, L. 1987, ApJ, 316, 127

Faber, S. M., and Gallagher, J. S. 1979, ARAA, 17, 135

Feigelson, E.D., Laurent-Muehleisen, S.A., Kollgaard, R.I., \& Fomalont, E. B. 1995, ApJ, 449, L149

Fomalont, E. B., Ebnerter, K. A., van Breugel,J. M., \& Ekers, R. D. 1989, ApJ, 346, L17

Geldzahler, B. J., \& Fomalont, E. B. 1984, AJ, 89, 1650

Gorgas J., Efstathiou  G. \& Aragon Salamanca, A. 1990, MNRAS, 245, 217

Hasinger, G., Burg, R., Giacconi, R., Hartner, G., Schmidt, M., Trumper, J., \& Zamorani, G. 1993, AA, 275, 1

Hibbard, J. E., Guhathakurta, P., van Gorkom, J. H., Schweizer, F. 1994, AJ, 107, 67

Jedrzejewski, R. I. 1987, MNRAS  226,  747

Kaneda, H., Tashiro, M., Ikebe, Y., Ishiski, Y., Kubo, H., Makishima, K., Ohashi, T., Saito, Y., Tabara, H., and Takahashi, T. 1995, ApJ, 453, L13

Killeen, N. E. B., Bicknell, G. V., \& Ekers, R. D. 1986a, ApJ, 302, 306

Killeen, N. E. B., Bicknell, G. V., \& Carter, D. 1986b, ApJ, 309, 45

Killeen, N. E. B., Bicknell, G. V., \& Ekers, R. D. 1988, ApJ, 325, 180

Kim, D.-W. 1989, ApJ, 346, 653

Kim, D.-W., Fabbiano, G., \& Trinchieri, G. 1992a, ApJS. 80, 645

Kim, D.-W., Fabbiano, G., \& Trinchieri, G. 1992b, ApJ. 393, 134

Kim, D.-W., \& Fabbiano, G., 1995, ApJ, 441, 182

Kim, D.-W., Fabbiano, G., Matsumoto, H, Koyama, K., \& Trinchieri, G. 1996, ApJ, 468, 175. 

Kley, W. \& Mathews, W. 1995, ApJ, 438, 100

Knapp, G. R., Turner, E. L., \& Cunniffe, P. E. 1985, AJ, 90, 454

Knapp, G. R., Guhathakurta, P., Kim, D.-W., \& Jura, M. 1989, ApJS, 70, 329

Kriss, G. A., Cioffi, D. F., \& Canizares, C. R. 1983, ApJ, 272, 439

Loewenstein, M., Mushotzky, R. F., Tamura, T., Ikebe, Y., Makishima, K., Matsushita, K., Awaki, H., \& Serlemitsos, P. J. 1994, ApJ, 436, L75.

Mackie, G., \& Fabbiano, G. 1997, AJ submitted

Malin, D. F., \& Carter, D. 1983, ApJ, 274, 534

Matsushita, K., Makishima, K., Awaki, H., Canizares, C. R., Fabian, A. C., Fukazawa, Y., Loewinstein, M.,  Matsumoto, H. Mihara, T., Mushotzky, R. F., Ohashi, T., Ricker, G. R., Serlemitsos, P. J., Tsuru, T., Tsusaka, Y., \& Yamazaki, T 1994, ApJ, 436, L41

Matsumoto, H., Koyama, K.,  Awaki, H., Tsuru, T., Loewenstein, M., and Matsushita, K. 1997, ApJ, 482, 133.

Pacholczyk, A. G. 1970, {\it Radio Astrophysics} (San Francisco: Freeman)

Pellegrini, S. \& Fabbiano, G. 1994, ApJ, 429, 105

Perley, R. A., Willis, A. G., \& Scott, J. S. 1979, Nature, 281, 437

Phillips, M. M., Jenkins, C. R., Dopita, M. A., Sadler, E. M. \& Binette, L.  1986, AJ, 91, 1062

Sage, L.J. \& Galletta, G., 1993, ApJ, 419, 544

Sarazin, C. L. 1988 {\it X-ray emissions from clusters of galaxies} (Cambridge: Cambridge University press)

Schweizer, F., 1980, ApJ, 237, 303

Schweizer, F., 1981, ApJ, 246, 722

Slee O. B., Sadler E. M., Reynolds, J. E. \& Ekers  R.D. 1994, MNRAS,  269, 928

Truemper, J. 1983, Adv. Space Res. 2, 241

Trinchieri, G. \& Fabbiano, G. 1991, ApJ, 382, 82

Trinchieri, G. \& di Serego Alighieri, S. 1991, AJ, 101, 1647

Trinchieri, G., Kim, D.-W., Fabbiano, G. \& Canizares, C. R. 1994, ApJ, 428, 555

Veron-Cetty, M. P. \&  Veron, P. 1986 , AAS, 66, 335

Wade, C. M. 1961, Pub. NRAO, 1, 99

Wiklind, T. \& Henkel, C. 1989, AA, 225, 1

Worrall, D. M., \& Birkinshaw, M. 1994, ApJ, 427, 134

\newpage
\bigskip
\centerline {Table 1}
\centerline {Basic parameters}
\medskip
\vbox{\tabskip=1em plus2em minus0.5em
\hrule\smallskip\hrule\smallskip
\halign to \hsize{#\hfil & #\hfil & #\hfil  \cr
                    &    NGC 1316     &     NGC 1317     \cr
\noalign{\smallskip\hrule\medskip}
RA (J2000)$^a$         &   3 22 41.6     &   3 22 44.7      \cr
DEC (J2000)$^a$        & -37 12 28       & -37  6  10       \cr
B$^o_T$ (mag)$^a$      &  9.40           &  11.81           \cr
D (Mpc)$^b$            &  27.2           &  27.2            \cr
D$_{25}$ (arcsec)$^a$  & 721             & 165              \cr
N$_H$ (cm$^{-2}$)$^c$  & 2.0 x 10$^{20}$ & 2.0 x 10$^{20}$  \cr
HRI Observed Date$^d$  & Jan. 14, 1994   & Jan. 14, 1994    \cr
HRI Exp time (sec)$^d$ & 11038           & 11038            \cr
HRI Observed Date$^e$  & Jul. 7-10, 1994 & Jul. 7-10, 1994  \cr
HRI Exp time (sec)$^e$ & 29403           & 29403            \cr
PSPC Observed Date     & Jan. 13-20, 1992& Jan. 13-20, 1992 \cr
PSPC Exp time (sec)    & 25500           & 25500            \cr
Log Fx (IPC) \esc $^f$ & 2.0 x 10$^{-12}$& $<$2.7 x 10$^{-13}$ \cr}
\smallskip\hrule\medskip}

\noindent {a. Right Ascension (RA), declination (DEC), total face-on B
magnitude (B$^o_T$), and major isophotal diameter measured at B = 25
magnitude arcsec\mtwo (D$_{25}$) taken from de Vaucouleurs et al.
1991 (RC3)}\\
\noindent {b. Distance from Fabbiano et al. 1992.}\\
\noindent {c. Galactic line of sight HI column density from Starks et
al. 1992}.\\
\noindent {d. Sequence number 600255n00}\\
\noindent {e. Sequence number 600255a01}\\
\noindent {f. IPC flux from Fabbiano et al. 1992. Fluxes were
estimated in a energy range of 0.2--4.0 keV for a Raymond-Smith model
with solar abundance, kT=1keV and line of sight $N_H$. The count
extraction radii are r=450\as\ for NGC 1316}.\\

\newpage
\bigskip
\centerline {Table 2}
\centerline {X-ray sources}
\medskip
\vbox{\tabskip=1em plus2em minus0.5em
\hrule\smallskip\hrule\smallskip
\halign to \hsize{
#\hfil &\hfil#\hfil &\hfil#     &\hfil#  & #\hfil   & \hfil#  & \hfil#
& \hfil# \cr
source &\hfil X  Y\hfil  &radius$^c$&offaxis&vignetting&net &error
&Fx$^e$ \ \cr
number &\hfil pixel\hfil &arcsec   &arcmin &correction&cnts&      &
$10^{-13}$ \cr
\noalign{\smallskip\hrule\medskip}
1   & 5599.86 3823.70 &   30 &  12.9 &   1.076 &     67.44   & 15.77 &  0.53 \cr
2   & 5075.38 4998.42 &   25 &  10.9 &   1.055 &     89.45$^d$&14.96 &  0.69 \cr
3   & 4468.47 4499.67 &   30 &   4.4 &   1.012 &     48.61   & 15.52 &  0.36 \cr
4   & 4104.82 3607.70 &   25 &   4.7 &   1.013 &     86.11   & 15.30 &  0.64 \cr
5$^a$&4076.98 4108.50 &  170 &   --- &    ---  &   2047.79   & 85.77 & 14.99 \cr
6$^b$&4009.78 4862.26 &   20 &   6.2 &   1.020 &    120.29   & 15.05 &  0.90 \cr
7   & 3842.10 4436.98 &   25 &   3.4 &   1.008 &     35.52   & 13.19 &  0.26 \cr
8   & 3702.10 3687.70 &   30 &   5.1 &   1.015 &     68.92   & 16.08 &  0.51 \cr
9   & 3296.79 3583.91 &   15 &   8.1 &   1.033 &     31.89   &  9.79 &  0.24 \cr
10  & 3041.20 3157.04 &   25 &  12.0 &   1.065 &     53.90   & 13.76 &  0.42 \cr
11  & 2959.06 4396.66 &   20 &   9.7 &   1.045 &     44.42   & 11.83 &  0.34 \cr
12  & 2823.38 3703.22 &   20 &  11.2 &   1.057 &    201.14$^d$&17.81 &  1.56 \cr}
\smallskip\hrule\medskip}

\noindent a. NGC 1316 \\
\noindent b. NGC 1317 \\
\noindent c. Background counts were extrated in annuli (r=200\as -
400\as\ for NGC 1316; r=60\as - 200\as\ for all other sources). \\
\noindent d. variable (see Table 3). \\
\noindent e. Fluxes were estimated in a energy range of 0.1--2.4 keV
for a Raymond-Smith model with solar abundance, $kT$ = 1 keV and line
of sight $N_H$. Vignetting correction was applied. 

\newpage
\bigskip
\centerline {Table 3}
\centerline {Variable Sources}
\medskip
\vbox{\tabskip=1em plus2em minus0.5em
\hrule\smallskip\hrule\smallskip
\halign to \hsize{
#\hfil &\hfil#\hfil  & \hfil #\hfil   & \hfil#  \cr
source &       Jan. 94   &    Jul. 94  &   difference  \cr
number &     rate \   err & rate \   err  &   $\sigma$   \cr
\noalign{\smallskip\hrule\medskip}
 2     &    -0.47 \  0.56&  3.22 \  0.46  &   5.09            \cr
12     &     7.81 \  1.04&  3.92 \  0.46  &   3.41            \cr}
\smallskip\hrule\medskip}

\noindent Rates and errors are in unit of counts in 1,000 seconds.

\newpage
TABLE4
\newpage
TABLE5

\newpage
\parskip=10pt
\centerline {Figure Captions}

Figure 1. The entire field of view of this HRI observation (NGC 1316
and NGC 1317). The X-ray contours are overlaid on the optical image
obtained from the Digital Sky Survey. X-ray image is binned with a
pixel size of 8 arcsec, background-subtracted and smoothed with a
Gaussian of $\sigma$ = 16 arcsec.  The octagonal shape indicates the
boundary of the HRI detector. The source numbers are ordered by an
increasing RA (see Table 2). RA and Dec are in J2000.

Figure 2. A close-up view of the X-ray image (with a pixel size of
2\as\ and a Gaussian \sig\ of 4\as). The contours indicate isophotes
at 5\% to 95\% of the peak with 8 steps. RA and Dec are in J2000.

Figure 3. (a) Radial distribution of X-ray counts extracted in
different angular sectors.  The filled circles are for the NE-SW
sector (PA=30-70\deg, 210-250\deg) and the open circles are for
NW-SE (PA=120-160\deg, 300-340\deg).  (b) The significance of count
differences between two angular sectors.

Figure 4. same as Figure 2 but with a pixel size of 1\as\ and a
Gaussian \sig\ of 2\as. The data obtained only in the July run are
used.  The contours indicate isophotes at 6\% to 85\% of the peak with
8 steps.

Figure 5. Radial distribution of X-ray counts determined with the raw,
unsmoothed image. Raw, background (determined at r=200--400\as), and
net counts are indicated by open squares, a solid line and filled
squares with error bars.

Figure 6. Radial profile of X-ray surface brightness and best fit
model prediction.  

Figure 7. (a) Deprojected density and (b) cooling time as a function
of radius. The density profile corresponding to $\beta$=0.51 ($n_e
\sim r^{-1.53}$) is shown as a dashed line.

Figure 8. Thermal gas pressure as a function of radius. The minimum
radio pressure along the jets is also plotted with filled circles (the
NW jet) and open circles (the SE jet).

Figure 9. Distribution of dust patches (from Schweizer 1980) overlaid
onto the X-ray image (same as Figure 2).

Figure 10. Distribution of ionized gas overlaid onto the X-ray image. The
continuum-subtracted \Ha\ + [NII] image was smoothed with a 2 pixel
(full width) gaussian filter and the X-ray image is the same as Figure 2.

Figure 11. Radio jet overlaid onto the X-ray image. The radio map was
taken from Geldzahler and Fomalont (1984). The X-ray image is the same
as (a) Figure 2 and (b) Figure 4.

\bye